\documentclass[aps,prl,twocolumn,superscriptaddress,amsfont,graphicx,nofootinbib,preprintnumbers]{revtex4}%
\usepackage{color,graphicx,epsfig}
\usepackage{ifpdf}
\usepackage{amsmath}
\usepackage{bm}
\usepackage{color}
\usepackage[english]{babel}
\usepackage{graphicx}%
\usepackage{amsfonts}%
\usepackage{amssymb}
\usepackage{braket}
\usepackage{hyperref}

\definecolor{nicered}{rgb}{0.7,0.1,0.1}
\definecolor{nicegreen}{rgb}{0.1,0.5,0.1}
\hypersetup{colorlinks,citecolor= nicegreen,linkcolor= nicered}

\newcommand{\slashed}{\slash \hspace{-0.19cm}}

\newcommand{\beq}{\begin{equation}}
\newcommand{\eeq}{\end{equation}}
\newcommand{\bea}{\begin{eqnarray}}
\newcommand{\eea}{\end{eqnarray}}

\definecolor{Red}{rgb}{1.,0.,0.}

\def\mysection#1{{{\bf #1}.~}}
\begin{document}

\def\LjubljanaFMF{Faculty of Mathematics and Physics, University of Ljubljana,
 Jadranska 19, 1000 Ljubljana, Slovenia }
\def\Cincy{Department of Physics, University of Cincinnati, Cincinnati, Ohio 45221,USA}
\def\LjubljanaIJS{Josef Stefan Institute, Jamova 39, 1000 Ljubljana, Slovenia}

%\preprint{}

\title{Flipping $t \bar t$ asymmetries at the Tevatron and the LHC}

\author{Jure Drobnak} 
\email[Electronic address:]{jure.drobnak@ijs.si} 
\affiliation{\LjubljanaIJS}

\author{Jernej F.\ Kamenik} 
\email[Electronic address:]{jernej.kamenik@ijs.si} 
\affiliation{\LjubljanaIJS}
\affiliation{\LjubljanaFMF}

\author{Jure Zupan} 
\email[Electronic address:]{jure.zupan@cern.ch} 
%\affiliation{\LjubljanaIJS}
%\affiliation{\LjubljanaFMF}
\affiliation{\Cincy}

\date{\today}
\begin{abstract}
We show that the charge asymmetry in $t\bar t$ production at the LHC, $A_C$, and the forward-backward asymmetry at the Tevatron, $A_{FB}$, are in general not tightly correlated. They can even have opposite signs, if the underlying new physics (NP) model is general enough. We demonstrate this using two examples of NP:  a light axigluon, and a vector that is a color octet and electroweak triplet. The small value of $A_C$ measured at the LHC is thus shown not to exclude a NP interpretation of the anomalously large $A_{FB}$ at the Tevatron. We identify two observables where significant NP effects are still expected at the Tevatron and the LHC, the $b\bar b$ production forward-backward asymmetry and spin polarizations of the pair-produced tops and anti-tops.
\end{abstract}

\maketitle

\mysection{Introduction} The forward-backward asymmetry (FBA) in $t\bar t$ production at the Tevatron has been measured by both the CDF~\cite{AFBCDF,AFBCDF1} and D\O~\cite{AFBD0} collaborations and found to be significantly larger than the standard model (SM) predictions. The na\"ive average of the inclusive FBA, 
adding the uncertainties in quadrature, is
\beq
A_{FB} = 0.187\pm 0.037\,,
\eeq
while the NLO QCD prediction~\cite{AFBCDF,MCNLO} including leading electroweak (EW) contributions~\cite{EW} is $A^{\rm SM}_{FB} = 0.07(2)$. Both CDF and D\O \, have also measured the FBA in bins of $m_{t\bar t}$ and $t$-$\bar t$ rapidity differences. Only CDF~\cite{AFBCDF,AFBCDF1}, however, unfolds to the partonic (``truth") level obtaining
\begin{subequations}
\begin{align}
A_{FB}^{\rm lo}\equiv A_{FB}(m_{t\bar t}<450~{\rm GeV}) &= 0.078 \pm 0.054\,,\\
A_{FB}^{\rm hi}\equiv A_{FB}(m_{t\bar t}>450~{\rm GeV}) &= 0.296 \pm 0.067\,,
\end{align}
\end{subequations}
to be compared with the SM (NLO QCD and EW) predictions $(A^{\rm lo}_{FB})^{\rm SM} = 0.05(1)$ and $(A^{\rm hi}_{FB})^{\rm SM}= 0.11(2)$~\cite{AFBCDF,MCNLO,EW}.

A related observable at the LHC is the charge asymmetry (CA) in $t\bar t$ production, $A_C$. In contrast to the FBA, the measurements of the CA at the LHC agree with the SM expectations. The average of ATLAS~\cite{ACATLAS} and CMS~\cite{ACCMS} results,
\beq\label{ACinclusive}
A_C = 0.001 \pm 0.014\,,
\eeq
agrees within errors with the SM prediction $A_C^{\rm SM} = 0.007(1)$~\cite{ACATLAS,MCNLO,EW}. Recently, the ATLAS collaboration also presented the first results for the CA binned in $m_{t\bar t}$~\cite{ATLASACmtt}
\begin{subequations}
\bea
A_C^{\rm lo}\equiv A_C(m_{t\bar t}<450~{\rm GeV}) &=& -0.053 \pm 0.088\,,\\
A_C^{\rm hi}\equiv A_C(m_{t\bar t}>450~{\rm GeV}) &=& -0.008 \pm 0.047\,,
\eea
\end{subequations}
in agreement with the corresponding SM predictions, $A_C^{\rm lo} = 0.002(2)$ and $A_C^{\rm hi}= 0.009(2)$~\cite{ACATLAS,MCNLO,EW}.

The CA measurements  pose a problem for new physics (NP) models addressing the FBA puzzle. Existing NP explanations of the anomalously large FBA predict  a non-negligible positive CA at the LHC in tension with data~\cite{ttreview,AguilarSaavedra:2011hz}. Should we then conclude that the observed FBA is not due to NP but a statistical fluctuation? Ideally, this would have been resolved with more precise measurements of the FBA. However, since the Tevatron stopped data taking this is no longer possible. The situation therefore needs to be clarified by the LHC.

In this Letter we derive several results relevant for LHC searches probing NP models that can explain the FBA. First of all, we show that sizable NP contributions to the FBA do not in general imply a deviation in the CA at the LHC.
We construct a simple modification of a widely discussed axigluon model which now predicts a large positive FBA, a small (or even negative) CA, and is consistent with all present measurements of the relevant $t\bar t$ production and other related collider observables. We also discuss a model with a light vector that is a color octet and an electroweak triplet.

Secondly, in order to obtain a nonzero FBA, the NP degrees of freedom need to couple to tops and light quarks. This means that inevitably there are  NP contributions to processes with light quarks or tops in the final state at the LHC. Furthermore, in models where there is little correlation between the CA and FBA, NP will couple with different strength to the different chiralities of quarks. A generic expectation is therefore that tops and antitops produced at the LHC will be partially polarized.

\mysection{General Considerations}
At the partonic level the FBA and CA are both due to the same charge asymmetric part of the $q\bar q \to t\bar t$ cross-section -- the terms that are proportional to the difference $\hat t-\hat u$ of the partonic Mandelstam variables (this was recently used to construct ``collider independent" asymmetries \cite{AguilarSaavedra:2012va}). A rigid positive correlation between the FBA and CA is then obtained in two cases: (i) either the NP contribution in $q\bar q \to t\bar t$ is due to only one light quark ($q=u$ or $d$), or (ii) the couplings to $u$ and $d$ are flavor universal. 

The correlation can be easily lost, if contributions of both $\bar u u$ and $\bar d d$ currents are significant and of opposite signs. This is because $p\bar p$ and $pp$ initial states have different valence structures. At large parton energy fraction $\tau = \hat s / s$, where $s$ is the collider energy squared, the ratio of parton luminosity functions $f(u\bar u)/f(d\bar d)$ at the Tevatron is roughly twice the one at the LHC, where the anti-quarks are non-valence. Both ratios decrease towards 1 at low $\tau$. Since at the LHC smaller values of $\tau$ are probed for the same $m_{t\bar t}$, this further enhances the difference between the Tevatron and the LHC. It is then possible to have simultaneously a large positive FBA and a small CA, if the partonic charge-asymmetry for $u\bar u \to t\bar t$ is positive, while it is negative for $d\bar d \to t\bar t$. 

%%%%%%%%%%%%%%%
\begin{figure}[t]
\centering{
\includegraphics[width=0.35\textwidth]{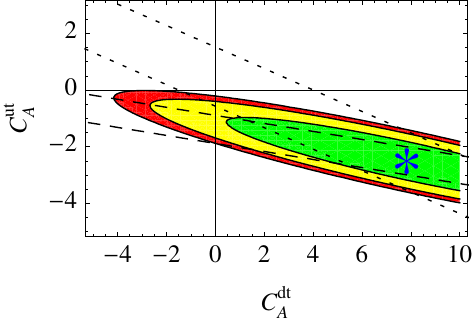}
}
\caption{\footnotesize The $1\sigma$ (green), $2\sigma$ (yellow) and $3\sigma$ (red) regions for a fit of EFT NP contributions in Eq. \eqref{eq:NPEFT} to the inclusive FBA (dashed) and CA (dotted) measurements. The best fit point where the central measured values of both CA and FBA are reproduced is marked with an asterisk.}
\label{fig:1}
\end{figure}
%%%%%%%%%%%%%%%

The above insight is easiest to check against data in the effective field theory (EFT) approach (we discuss on-shell models below). The FBA and CA can be generated from the interference of the leading order SM amplitudes and NP contributions. At ${\mathcal O}(\alpha_S \Lambda^{-2})$ there are only two relevant dimension 6 NP operators 
\beq
\label{eq:NPEFT}
\mathcal L = \mathcal L_{\rm SM} + \sum_{q=u,d}\frac{C_A^{qt}}{\Lambda^2} (\bar q \gamma^\mu \gamma_5 q) ( \bar t \gamma_\mu \gamma_5 t)\,,
\eeq
where $\Lambda$ is the NP scale, and $C_A^{qt}$ the NP Wilson coefficients. Note that at $\mathcal O(\alpha_s \Lambda^{-2})$ the NP operators in \eqref{eq:NPEFT} do not affect the forward-backward symmetric $t\bar t$ cross-section, while they do generate shifts in inclusive FBA and CA
\beq\label{numerics-AFBAC}
\begin{split}
\Delta A_{FB}&= -10\%\times \big(0.84 C_A^{ut} + 0.12 C_A^{dt}\big)\big({1\rm TeV}/{\Lambda}\big)^2,\\
\Delta A_{C}&= -1\% \times\big(1.4 C_A^{ut} + 0.52  C_A^{dt}\big)\big({1\rm TeV}/{\Lambda}\big)^2.
\end{split}
\eeq
A large FBA and small or negative CA are possible, if $C_A^{dt}$ and $C_A^{ut}$ have opposite signs and $|C_A^{dt}| \gtrsim |C_A^{ut}|$.  Both CA and FBA measurements can then be accommodated as shown in Fig.~\ref{fig:1}.
 A fairly large value of $C_A^{dt}$ is required or a correspondingly low NP scale $\Lambda$, so the EFT description is likely not valid. However,  the generic requirements -- a large coupling to $\bar d d$ quark currents, and $\bar u u$ contributions of the opposite sign -- are expected to apply also to on-shell NP models.

In passing we also mention a second effect that breaks the correlation between the FBA and CA. The luminosity functions are falling faster with $\hat s\sim m_{t \bar t}$ at the Tevatron than at the LHC. Therefore it would be theoretically possible that at the LHC a small inclusive CA  is due to a cancellation between a positive CA at low $m_{t\bar t}$ and a large negative CA at large $m_{t\bar t}$ (at the Tevatron the positive contribution would dominate). This mechanism would yield FBA and CA falling with $m_{t\bar t}$ in contrast to observations.

We next consider explicit on-shell NP models, and discuss in turn the asymmetric axigluon, and the electroweak triplet of color octet vectors. These models are representative of scenarios where the cancellation in the CA can occur.%: one requires both  $u-t$ and $d-t$ couplings to NP.

\mysection{Asymmetric Axigluon}
The model is a simple modification of the light axigluon model originally introduced by Tavares and Schmaltz~\cite{Schmalz} (see also \cite{Frampton:2009rk}). An $SU(3)_L\times SU(3)_R$ gauge symmetry is broken spontaneously via a bifundamental scalar $\phi$ to the diagonal $SU(3)_{\rm color}$.  The SM fermions are charged under the full gauge group and transform as $Q=(3,1)$ and $U,D=(1,3)$. Additional heavy vector-like fermions are integrated out at the scale $\Lambda$ inducing dimension five and six interactions between the light (SM-like) fermions and the axigluon~\cite{Schmalz}. 
The relevant effective Lagrangian at scales $\braket{\phi} \lesssim \mu \lesssim \Lambda$ is then
\bea
\mathcal L &=& -\frac{1}{4} (F_L^a)^2 -\frac{1}{4} (F_R^a)^2 + \bar Q i\slashed D Q + \bar U i\slashed D U + \bar D i\slashed D D \nonumber\\
&&+ \frac{1}{\Lambda^2} \left[ \lambda_Q^2 \overline{(\phi^\dagger Q)} i\slashed D (\phi^\dagger Q ) + \lambda_U^2 \overline{(\phi U)} i\slashed D (\phi U ) \right. \nonumber\\
&&\left. + \lambda_D^2 \overline{(\phi D)} i\slashed D (\phi D )  \right] + \mathcal L_{\rm Yuk} + \mathcal L_{\phi}\,,
\eea
where $\mathcal L_{\rm Yuk}$ are the SM Yukawa terms and $\mathcal L_{\phi}$ contains the kinetic and potential terms for the scalar field $\phi$. 

To decorrelate FBA and CA the only necessary modification of the original construction~\cite{Schmalz} is to allow for sizable parity breaking in the new fermionic sector. As a consequence $\lambda_Q\ne \lambda_U\ne\lambda_D$. 
For notational simplicity we keep $\lambda_{Q,U,D}$ flavor universal and the gauge interactions left-right symmetric, $g_L=g_R\equiv g$ (we will relax both assumptions below). 
After $SU(3)_L\times SU(3)_R\to SU(3)_{\rm color}$ breaking,
diagonalizing the gauge boson mass matrix, and rescaling the fermion fields, one obtains a new effective Lagrangian at the EW scale
\bea
\mathcal L &=& -\frac{1}{4} (G_{\mu\nu}^a)^2 -\frac{1}{4} (\tilde G_{\mu\nu}^a)^2 + \frac{\tilde m^2}{2} \tilde A_\mu^2 + \bar Q (i\slashed D - \tilde g_Q \slashed {\tilde A}) Q \nonumber\\
&&+ \bar U (i\slashed D + \tilde g_U \slashed {\tilde A}) U + \bar D (i\slashed D + \tilde g_D \slashed {\tilde A}) D + \ldots\,,
\label{axilagr}
\eea
where $G_{\mu\nu}^a$ is the gluon field strength, $\tilde A_\mu \equiv (A_{R\mu}-A_{L\mu})/\sqrt 2$ is the axigluon field,  $\tilde G^a_{\mu\nu}$ the axigluon field strength, and the covariant derivative $D_\mu$ now contains only  the gluon field. 
If the initial $SU(3)_R\times SU(3)_L$ gauge interactions are left-right symmetric the axigluon couplings to fermions are bounded by $|\tilde g_{Q,D,U}|\leq g_s$\,. In the general case where $g_R\ne g_L$, the couplings $\tilde g_{Q,D,U}$ can be arbitrarily large (up to the perturbative limit).

The partial decay width for the axigluon decaying to $q\bar q$ pairs is
\beq
\frac{\Gamma_q}{\tilde m} = \frac{\sqrt{1-4r_q}}{48\pi} \left[ (1-r_q) ({\tilde g_Q}^2 + {\tilde g_{U,D}}^2) + 6 r_q \tilde g_{U,D} \tilde g_Q  \right]\,,
\eeq
where $r_q = m_q^2/\tilde m^2$\, and the $\tilde g_{U,D}$ couplings apply for decays to up- and down-type quarks, respectively.  The axigluon lighter than $2m_t$  thus has a total decay width of
\beq
{\Gamma} \simeq \frac{\tilde m}{48\pi} \left[ 5 {\tilde g_Q}^2 + 2 {\tilde g_{U}}^2 + 3 {\tilde g_{D}}^2 \right]\,.
\eeq
For $\tilde g_i \sim g_s \sim \mathcal O(1)$,  the  axigluon decay width is sizeable, $\Gamma \sim 0.1 \tilde m$.

%%%%%%%%%%%%%%%
\begin{figure}[t]
\includegraphics[width=0.35\textwidth]{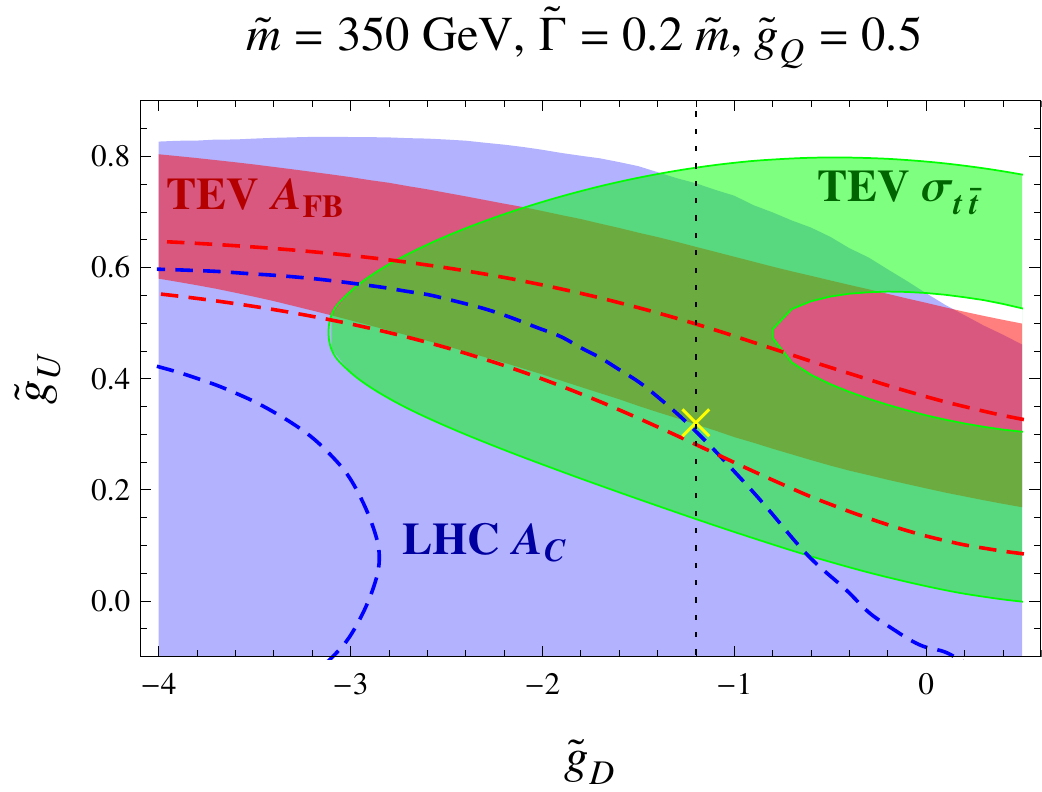}
\caption{\footnotesize Predictions for the relevant $t\bar t$ observables in the axigluon model \eqref{axilagr} with mass $\tilde m=350$ GeV, decay width $\Gamma=0.2 \tilde m$, and $\tilde g_Q=0.5$, when varying $\tilde g_D, \tilde g_U$,  compared to the $1\sigma$ constraints from inclusive $A_C$ (dashed blue lines),  from $A_C^{\rm hi}$ (blue band),  inclusive $A_{FB}$ (dashed red lines), 
 from $A_{FB}^{\rm hi}$ (red band), and inclusive $\sigma_{t\bar t}$ (green band)~\cite{sigmaTEV}. The vertical dotted line is the boundary of the left-right symmetric model $\tilde g_D=-g_s$. Yellow cross denotes the chosen benchmark point (see text for details).
 }
\label{fig:modelfit}
\end{figure}
%%%%%%%%%%%%%%%

The agreement with $A_C$, $A_{FB}$, and the inclusive $t\bar t$ cross section ($\sigma_{t\bar t}$) at the Tevatron~\cite{sigmaTEV,sigmaTEVSM} is shown in Fig. \ref{fig:modelfit} for an axigluon mass $\tilde m=350$ GeV and decay width fixed to $\Gamma=0.2\tilde m$ for simplicity (this decay width is saturated for $|\tilde g_D|\simeq 3$, otherwise flavor non-universal couplings to $b_R$,$s_R$ (larger than to $d_R$) are implicitly assumed). The predictions are made using {\tt FeynRules1.5.48} \cite{Christensen:2008py}, {\tt Madgraph5.1.3.30} \cite{Alwall:2011uj} with {\tt Pythia6.425} \cite{Sjostrand:2006za} + {\tt PGS4} \cite{PGS4} pipeline. Scanning over $\tilde g_{Q,U,D}$ the best fit region is obtained for $\tilde g_Q \simeq \tilde g_U \simeq 0.5$ and $\tilde g_D \simeq -2$, where the NP predictions are well within the $1\sigma$ experimental regions. As a benchmark we choose a point on the boundary of the $\tilde g_D$ values in the left-right symmetric axigluon model, $\tilde g_Q = 0.5, \tilde g_U = 0.32, \tilde g_D = -1.2$.
For this point we obtain the central values for inclusive $A_{FB}= 0.16$ and $A_C=0.015$, while in the high $m_{t\bar t}>450$ GeV region we predict $A_{FB}^{\rm hi}=0.23$ and $A_C^{\rm hi}=0.019$.

\begin{figure}[t]
\includegraphics[width=0.35\textwidth]{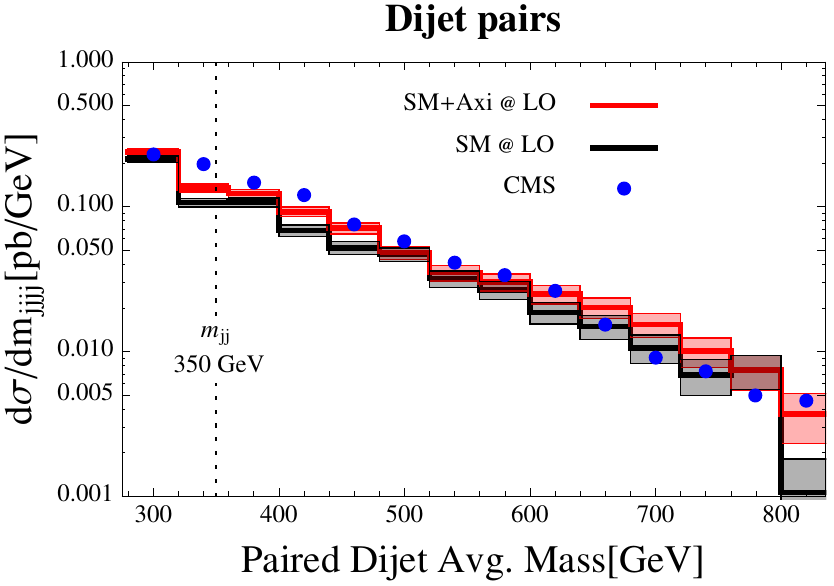}
\caption{\footnotesize Comparison of the CMS data on paired dijet production~\cite{CMS-paired-dijets} (blue points), the SM LO prediction (black) and the axigluon model benchmark point from Fig. \ref{fig:modelfit} (red). The bands of the two theory curves are estimates of statistical Monte Carlo errors and are not representative of SM theory errors.}
\label{fig:paired-dijets}
\end{figure}

We have checked that the benchmark satisfies all the remaining LHC and Tevatron constraints. The $350$ GeV axigluon is below the $t\bar t$ threshold and does not produce a resonance in the differential $d\sigma_{t\bar t}/d m_{t\bar t}$ distribution which is then in good agreement with the measurements. The $d\sigma_{t\bar t}/d m_{t\bar t}$ spectrum would be an important constraint, though, for a heavier axigluon with non-universal  $\tilde g_{Q}\ne \tilde g_{U}\ne \tilde g_{D}$ resulting in vectorial couplings of the axigluon to the SM quarks.
The bump hunting and angular correlations measurements in dijet production at the LHC and the Tevatron are not yet sensitive to the axigluon with the benchmark point couplings (even if the decay width is smaller, e.g. $\Gamma=0.1 \tilde m$). More constraining is the CMS resonance search in paired dijets  \cite{CMS-paired-dijets}. A qualitative comparison of  the CMS data with our simulated LO SM and axigluon model predictions is shown in Fig. \ref{fig:paired-dijets}, where we have assumed that the axigluon decays to two jets with a $100\%$ branching fraction. For smaller decay widths, e.g., already for $\Gamma=0.1 \tilde m$, there would be an observable resonance peak in the distribution, which is excluded. A decay width of $\Gamma=0.2 \tilde m$ implies that at our benchmark point the axigluon has large couplings to either $s_R$, $b_R$ or both, for instance $\tilde g_D(s_R)=\tilde g_D(b_R)=-3.7$, and can be searched for using the $b\bar b$ forward-backward asymmetry as we show below.

%%%%%%%%%%%%%%%
%\begin{figure}[t]
%\includegraphics[width=0.4\textwidth]{figs/mtt_IVoFW_450_45}
%\caption{\footnotesize The $d\sigma/dm_{t\bar t}$ predictions for asymmetric axigluon (kodrlajsasta) and colored EW triplets (solid) benchmark models, compared to LO SM (dotted) and CDF measurement (blue $1\sigma$ band) \cite{CDFmtt}. {\bf JZ: only the axigluon, and appropriately changed caption.}}
%\label{fig:ttbar-spectrum}
%\end{figure}
%%%%%%%%%%%%%%%

%%%%%%%%%%%%%%%
%%%%%%%%%%%%%%%

\begin{figure}[t]
\includegraphics[width=0.35\textwidth]{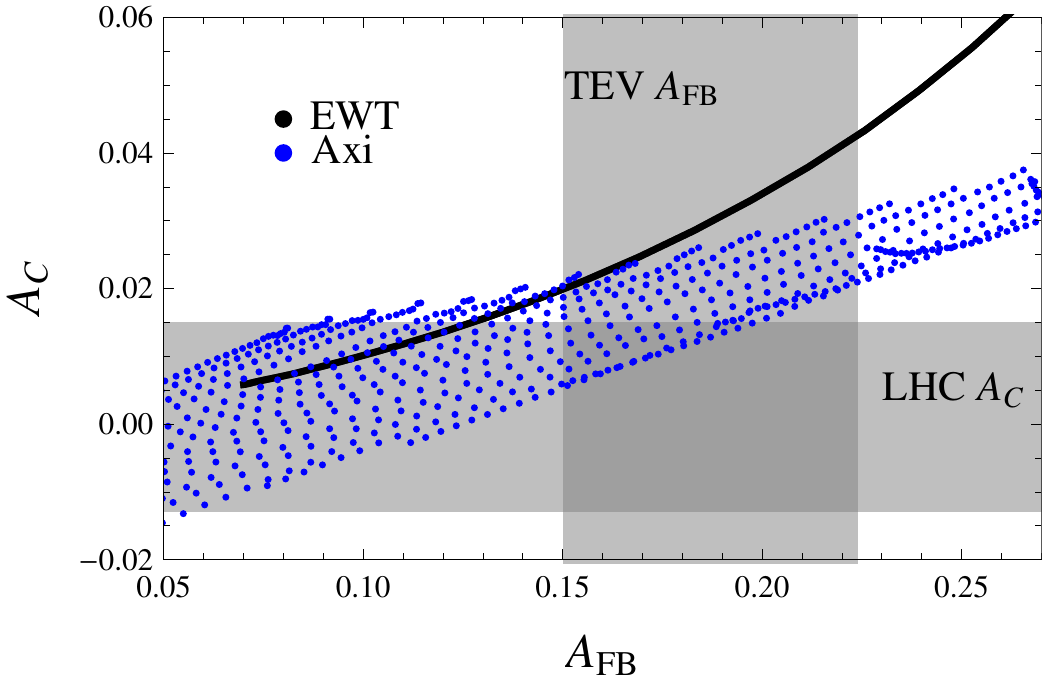}
\caption{\footnotesize The correlation between $A_{FB}$ and $A_C$ in two models,  the colored EWT model (black) and the asymmetric axigluon model (blue), the later allowing for much larger spread in $A_C$ at fixed $A_{FB}$. The scan is over the range of $\tilde g_{U,D}$ in Fig. \ref{fig:modelfit} for the axigluon and over $\eta_0\in [0,2.6]$ for the EWT, while $\tilde m = m_{V}=350$ GeV and $\Gamma =70$~GeV in both models. Gray bands represent the experimental $1\sigma$ regions for the two observables.}
\label{fig:scatter}
\end{figure}

\mysection{Electroweak Triplets} 
The cancellation between the $u\bar u$ and $d\bar d$ contributions is automatic for NP resonances that are electroweak triplets (EWT).
As an example we consider a color octet EWT vector (model IV$_{\rm o}$ in \cite{Grinstein:2011dz}) with the interaction Lagrangian 
\beq\label{color-EWtriplet}
{\cal L}=\eta_0 \bar Q_L T^a \tau^i \slashed V^{a,i} Q_L+\cdots,
\eeq
where the $SU(3)_{\rm color}$ and $SU(2)_L$ generators are normalized to ${\rm Tr}(T^a T^b)={\rm Tr}(\tau^a\tau^b)=\delta_{ab}/2$.
The  $t\bar t$ production asymmetry is due to the $s-$channel exchange of the charge neutral $V^{a,3}$ resonances. They couple to $u_L$ and $d_L$ with opposite signs, which leads to a natural suppression of $\Delta A_C$. Since the relative sizes and signs of these couplings are fixed by gauge symmetry, contributions to the CA and FBA are tightly correlated. This is illustrated in Fig.~\ref{fig:scatter} where we vary $\eta_0\in [0,2.6]$, but fix $m_{V}=350$ GeV and $\Gamma=0.2 m_{V} $ for ease of comparison with the asymmetric axigluon model.
A more comprehensive scan over $\eta_0, m_V$ and $\Gamma$   
%and also $\eta_1 \in [...,...]$ (breaking flavor universality between the couplings to the first and the third generation quarks) 
comparing the predictions with $A_{FB}^{\rm hi}$, $A_{FB}$, $A_C^{\rm hi}$, $A_C$ and $d\sigma_{t\bar t}/dm_{t\bar t}$, reveals, however, that it is not possible to simultaneously satisfy all the above constraints within $1\sigma$. The main reason is that accommodating both the FBA and the CA simultaneously, requires couplings to $\bar d d$ currents much larger than to $\bar u u$ (c.f. Eq.~\eqref{numerics-AFBAC}), something not allowed by the gauge symmetric structure of the model.  Consequently, at present the model is disfavored. It could become viable again, if the present tension among the observed FBA and CA values is somewhat reduced by future more precise measurements. 

%%%%%%%%%%%%%%%
%%%%%%%%%%%%%%%%%%%%%%%%%%%%%%
\begin{figure}[t]
\includegraphics[width=0.35\textwidth]{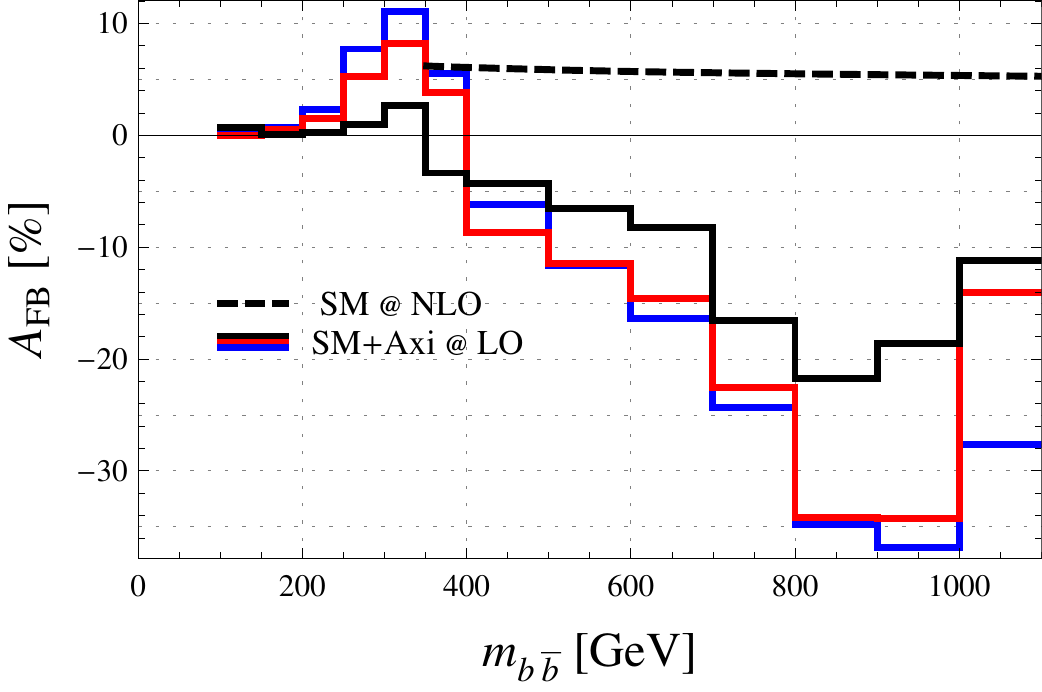}
\caption{\footnotesize The $b\bar{b}$ forward-backward asymmetry at the Tevatron as predicted by the  axigluon model \eqref{axilagr} with the benchmark parameter set from 
Fig.~\ref{fig:modelfit} (black) and with additional flavor symmetry breaking $\tilde g_D(s_R)=\tilde g_D(b_R)=-3.7$ (red) or  $\tilde g_D(b_R)=-5.1$ (blue).}
\label{fig:bbAsym}
\end{figure}
%%%%%%%%%%%%%%%

\mysection{Discovery observables} 
Finally, let us discuss the observables with the help of which the above NP models could be discovered at the LHC. Any NP model affecting the FBA and CA needs to couple to light quarks. More precise measurement of dijet production or paired dijet production could 
therefore reveal irregularities, though, as we have shown above, this could be challenging for broad resonances. 
 
There are also several other common features of the models in which $A_{C}$ is suppressed. For instance, to have a large cancellation between $u\bar u$ and $d\bar d$ contributions, the  couplings to down quarks are necessarily enhanced, c.f. Eq. \eqref{numerics-AFBAC}. This then generically implies a significant effect in the $b\bar b$ forward-backward asymmetry at the Tevatron (though precise predictions are model dependent). In the asymmetric light axigluon model, for  example, the couplings to $b_R$ and $s_R$ should be even further enhanced to make the axigluon broad. In Fig.~\ref{fig:bbAsym} we show three representative cases. The first one is the benchmark point in Fig.~\ref{fig:modelfit} with flavor universal couplings $\tilde g_D=-1.2$ for $d_R, s_R$ and $b_R$ (black line), for which the decay width would be small unless channels besides $q\bar q$ are open. The blue and red lines denote flavor nonuniversal choices where the couplings are changed either to $\nolinebreak{\tilde g_D(s_R)=\tilde g_D(b_R)=-3.7}$ or to $\tilde g_D(b_R)=-5.1$, respectively, while the other parameters are left the same (these choices give $\Gamma=0.2\tilde m$). The CDF sensitivity for $m_{b\bar b}>130$ GeV is $2.6\%$~\cite{CDFbbartalk}. The largest deviations appear for  $m_{b\bar b}\gtrsim 400$ GeV (i.e. above the resonance mass), thus the extension of measurements to  higher $b\bar b$ pair masses is highly desirable.

Another generic property of the models that will suppress the CA is that they contain new chiral couplings to quarks. (To suppress $A_C$, couplings to $u$ and $d$ need to be different. Purely vector or axial couplings are possible only if there is a fine-tuned electroweak symmetry breaking of the couplings to $Q_L$ with the sizes of $u_R$ and $d_L$ couplings.) This means that the $t$ and $\bar t$ produced through NP interactions will be polarized.  For the asymmetric axigluon benchmark point we obtain the top polarization fractions $B^{\rm TEV}_{\rm beam} \simeq B^{\rm TEV}_{\rm off-diagonal} \simeq 13\%$ and $B^{\rm TEV}_{\rm helicity} \simeq 7\%$ at the Tevatron, where the numbers refer to different choices of the top spin quantization axis (c.f.~\cite{Fajfer:2012si} and references therein). At the LHC the effects are significantly diluted. We find that only the helicity axis polarization, $B^{\rm LHC7}_{\rm helicity} \simeq 2\%$, is larger than a percent.

\mysection{Conclusions} We have shown that the FBA in $t\bar t$ production at the Tevatron can in general be large, $\mathcal O(0.2)$, and the CA at the LHC remain small $\lesssim \mathcal O(1\%)$ (or even of the opposite sign). The reason is that the $u\bar u $ and $d\bar d$ contents of $p\bar p$ and $p p$ initial states are different.  For the CA to be small, the coupling of NP to $\bar dd$ currents needs to be large and of opposite sign to the coupling of NP to $\bar u u$. We have shown this using an effective theory analysis as well as for two explicit on-shell models: a light asymmetric axigluon model and a model with colored electroweak triplet vectors.  The asymmetric light axigluon gives a good description of the present data (for both $A_C$ and $A_{FB}$ as well as other collider constraints), and predicts generically a large $b\bar b$ forward-backward asymmetry and observable top polarization in $t\bar t$ production. 

%\begin{acknowledgments}
\mysection{Acknowledgements}
We thank T. Mede for sharing computing resources with us. J.D. thanks C. Duhr and O. Mattelaer for enlightening discussion as well as the Lawrence Berkeley National Laboratory and University of Cincinnati, where part of this work was completed for their hospitality. This work was supported in part  by the Slovenian Research Agency. 
%\end{acknowledgments}

\end{document}